\documentclass[twocolumn,pdflatex]{aastex631}

\usepackage{here}
\usepackage{natbib}
\usepackage{bm}

\shorttitle{GRB Emission with Anisotropic Electron Distribution}
\shortauthors{Goto \& Asano}

\graphicspath{{./}{figures/}}

\begin{document}

\title{GRB Prompt Emission with Anisotropic Electron Distribution}

\author[0000-0003-2825-8903]{Ryota Goto}

\author[0000-0001-9064-160X]{Katsuaki Asano}
\affiliation{Institute for Cosmic Ray Research, The University of Tokyo,
5-1-5 Kashiwanoha, Kashiwa, Chiba 277-8582, Japan}

\begin{abstract}
The typical spectrum of the prompt emission of gamma-ray bursts (GRBs) indicates that the electron cooling is suppressed in spite of the strong magnetic field in the standard synchrotron model.
Recent Particle-in-Cell simulations show that the particle acceleration by magnetic reconnection in a magnetically dominated plasma can lead to small pitch angles especially in low-energy region.
Such a small pitch angle prevents electrons from cooling via synchrotron radiation.
In this paper, taking into account the effects of the synchrotron cooling and the adiabatic cooling, we numerically calculate the synchrotron spectra with anisotropic electron distributions.
If we require a Poynting flux larger than $10^{50}~\mbox{erg}~\mbox{s}^{-1}$
as the model is motivated by magnetic reconnection,
the bulk Lorentz factor of $\sim 1000$
and the electron minimum Lorentz factor of $\gamma_{\rm min}\sim 10^4$ are required
to reproduce the typical GRB spectrum.
\end{abstract}

\section{Introduction} \label{sec:intro}
The prompt emission of gamma-ray bursts (GRBs) is considered to be radiation from relativistic jets \citep[for reviews, see, e.g.,][]{2015PhR...561....1K,2015AdAst2015E..22P}.
The GRB prompt spectra are usually well fitted by the Band function \citep{1993ApJ...413..281B},
which is smoothly-broken power-law with the low and high energy spectral indexes, $\alpha$ (typically $\sim-1$) and $\beta$ ($\sim-2.2$), respectively, and the peak energy of the $\nu F_{\nu}$ spectrum $E_{\mathrm{p}}$ \citep[typically several hundreds $\mathrm{keV}$, see,][]{2000ApJS..126...19P,2006ApJS..166..298K,2011A&A...530A..21N,2014ApJS..211...12G}.
One of the promising models to explain the typical GRB spectrum is synchrotron radiation from electrons injected with a power-law energy distribution with a low-energy cutoff \citep{1996ApJ...466..768T,2000ApJ...543..722L}.
But there exist several problems in this standard synchrotron model.
Representative one of them is the synchrotron fast cooling problem \citep{2000MNRAS.313L...1G}.
Given a magnetic field consistent with $E_{\mathrm{p}}$,
electrons promptly lose energy via synchrotron radiation during the dynamical timescale,
and photons emitted from such cooled electrons contribute to the low-energy spectrum.
As a result, the predicted spectral index becomes $\alpha=-1.5$,
which is significantly softer than the typical index $\alpha\sim-1$.

Several alternative models to reconcile the observed hard spectra have been proposed.
The photosphere emission model is the most promising one \citep{2005ApJ...628..847R,2010MNRAS.407.1033B,2010ApJ...725.1137L,2011ApJ...732...49P}.
\citet{2010MNRAS.407.1033B} and \citet{2010ApJ...725.1137L} reproduce high energy power law component ($\beta\sim-2.2$) by comptonization of thermal photons emitted by subphotospheric dissipation.
Although the simple Planck spectrum implies a too hard spectrum ($\alpha=+1$),
the Doppler-boosted emission components from different angles with respect to the line of sight can soften the spectrum
as $\alpha \sim +0.4$ \citep{2010MNRAS.407.1033B} or $\alpha \sim -1$ \citep{2011ApJ...732...49P}.
\citet{2017MNRAS.468.3202B} proposed the magnetically-dominated jet model with the photospheric emission.
The model reproduces the observed spectrum with $\alpha\sim-1$ by superimposed subdominant synchrotron radiation
on the dominant thermal radiation.

\citet{2001A&A...372.1071D} and \citet{2011A&A...526A.110D} suggested that if the synchrotron self-Compton (SSC) cooling rate in the Klein-Nishina regime exceeds the synchrotron cooling rate, a hard synchrotron spectrum can be reproduced.
However, the SSC dominance in the GeV band is not confirmed in major fraction of GRBs observed with Fermi LAT \citep{2011ApJ...730..141Z,2013ApJS..209...11A}.
If the SSC emission in GeV--TeV range is absorbed in the source via electron--positron pair creation,
the electromagnetic cascade emission makes the low-energy spectrum soft significantly
\citep{2011ApJ...739..103A,2012ApJ...757..115A}.

The SSC emission in 0.1--1 MeV \citep{2004MNRAS.352L..35S,2009A&A...498..677B} is another option
to reproduce the GRB spectrum.
This SSC model predicts strong optical and GeV-TeV emissions \citep{2009MNRAS.393.1107P},
such a signature has not been established as a consensus in the majority of GRBs \citep{2011ApJ...730..141Z,2013ApJS..209...11A}.

The jitter radiation \citep{2000ApJ...540..704M} is an emission process in a small-scale magnetic turbulence.
This process leads to a harder spectrum ($\alpha=0$) than synchrotron emission.
But in the fast cooling regime, the jitter radiation results in $\alpha=-1.5$.
Even in this model, the suppression of the electron cooling is required.

In \citet{2021NatCo..12.4040R}, the observed evolutions of the spectral photon index and the flux density in the steep decay phase of the early afterglow, which is considered as a tail emission of a prompt emission, suggest that adiabatic cooling is the dominant cooling process.
Therefore, the suppression of the radiative cooling for low-energy particles
seems essential for the GRB emission process.
The proton synchrotron model \citep{2020A&A...636A..82G} is one of such models.
Alternatively, a continuous electron acceleration by turbulence
leads to the balance between acceleration and synchrotron cooling,
which produces a hard electron distribution \citep{2009ApJ...705.1714A,2015MNRAS.454.2242A,2017ApJ...846L..28X,2018MNRAS.476.1785B}.

If the decay timescale of the magnetic field is comparable
to the cooling timescale,
the inefficient cooling can produce the required spectral shape
\citep{2006ApJ...653..454P,2011A&A...526A.110D,2013ApJ...769...69B}.
\citet{2014NatPh..10..351U} calculated spectra of instantaneous emissivity after magnetic field decays and obtained hard spectra with $\alpha\sim-1.2$. However, in most of GRB observations, spectra are obtained by integrating photons for significantly longer timescale ($\gtrsim 1\mathrm{s}$) than the variability timescale. In this paper, we focus on the time-integrated spectra.

Recent Particle-in-Cell (PIC) simulations \citep{2018JPlPh..84f7201P,2019ApJ...886..122C} show that magnetic reconnection induced by turbulence in a magnetically dominated plasma produces anisotropic distribution of accelerated particles. 
The acceleration process in their simulations are divided into two:
the prompt acceleration along the guiding magnetic field at the reconnection site
and the succeeding turbulence acceleration similar to the 2'nd order Fermi acceleration.
The first prompt acceleration injects relatively low-energy electrons with small pitch angles.
High-energy electrons are accelerated by turbulence, which makes electrons isotropic.
In such cases, the synchrotron emission from low-energy electrons with small pitch angles
is suppressed.
The spectral hardening by such an anisotropic electron distribution has been
pointed out by several authors \citep{2000ApJ...543..722L,2018ApJ...864L..16Y}.
This possibility is discussed in the context of the SSC dominance in blazars \citep{2021MNRAS.506...38S}.

In this paper, we discuss the GRB spectrum with anisotropic distributions as suggested by \citet{2019ApJ...886..122C}.
The small pitch angle for low-energy electrons
suppresses the electron cooling even in a strong magnetic field.
The effect of the energy dependence of the anisotropy is not trivial for the synchrotron spectral shape.
We numerically solve the temporal evolutions of electron energy and pitch angle distributions,
which are essential to compare the time-integrated spectrum with observations.
We take into account adiabatic cooling and decrease of the magnetic field due to the GRB jet expansion.

The structure of this paper is as follows.
In \S \ref{sec:method}, we explain our calculation method of synchrotron spectra taking into account anisotropic electron distributions.
In \S \ref{sec:result}, we show the numerical calculation results of the spectra. 
and discuss the required parameter regions.
In \S \ref{sec:sum}, we summarize our results and discuss the validity of the anisotropic model.

\section{Method} \label{sec:method}

We consider a conical jet with a constant bulk Lorentz factor $\Gamma$ and an injection of electrons at a radius $R=R_0$.
We numerically solve the temporal evolutions of the electron energy and pitch angle distributions in the jet comoving frame. The comoving volume expands with its propagation, so we take into account adiabatic cooling and decrease of the magnetic field as well as synchrotron cooling.
Integrating photons emitted from those electrons, the photon spectrum is obtained.
Based on this time-integrated spectrum, we discuss the effects of the anisotropic distribution.

\subsection{\bf{Evolution of the Electron Distribution}}

The injection spectrum is written as follows,
\begin{eqnarray}
\label{eq:initial}
N(\gamma,\mu,t=0)&=&C\gamma^{-p}f(\gamma,\mu)\ \nonumber\\
& &(\gamma_{\rm{min}}<\gamma<\gamma_{\rm{max}}),
\end{eqnarray}
where $C$ is a normalization constant, $N(\gamma,\mu,t)$ is the electron number per $\gamma$ (the Lorentz factor of electrons) per $\mu=\cos\psi$ (the pitch angle cosine) in the comoving frame, and $f(\gamma,\mu)$ expresses the pitch angle distribution, which generally depends on $\gamma$.
We assume a power-law energy distribution with a low energy cutoff $\gamma_{\mathrm{min}}$, which has been conventionally assumed in the interpretation of the GRB emission \citep{1996ApJ...466..768T}.
We fix the electron spectral index $p=2.5$ to reproduce the observed high-energy spectral index $\beta\sim-2.2$ ($\beta \simeq -(p+2)/2$ in the fast cooling synchrotron spectrum), while we discuss cases with different $p$ values in \S \ref{sec:pitch}.

The time evolution of the electron distribution in the comoving frame
is given by the following continuity equation
in the $(\gamma,\mu)$-space,
\begin{eqnarray}
\label{eq:continuity}
\frac{\partial}{\partial t}N(\gamma,\mu,t)
+\frac{\partial}{\partial \gamma}\{(\dot{\gamma}_{\rm syn}+\dot{\gamma}_{\rm adi})N(\gamma,\mu,t)\}\nonumber\\
+\frac{\partial}{\partial \mu}\{(\dot{\mu}_{\rm syn}+\dot{\mu}_{\rm adi})N(\gamma,\mu,t)\}=0,
\end{eqnarray}
where $\dot{\gamma}$ and $\dot{\mu}$ are the energy and pitch angle changing rates, respectively.
The subscripts ``$\mathrm{syn}$'' and ``$\mathrm{adi}$'' mean the synchrotron and adiabatic cooling effects, respectively.
Here, we neglect the SSC cooling, which depends on the electron luminosity or normalization $C$,
and the magnetic field.
Thus, the discussion on the photon spectral shape in this paper does not depend on the normalization $C$.
If we consider a magnetically dominated jet, which is likely for the magnetic reconnection model,
the negligible contribution of the SSC cooling may be adequate.
For simplicity, we also neglect the pitch angle scattering by turbulence.
We will discuss the influence of the pitch angle scattering in \S \ref{sec:sum}.

The synchrotron energy changing rate \citep{1979rpa..book.....R} is 
\begin{eqnarray}
\label{eq:synenergy}
\dot{\gamma}_{\rm syn}=-\frac{2e^4B^2(\gamma^2-1)(1-\mu^2)}{3m_{\mathrm{e}}^3c^5},
\end{eqnarray}
where $B$ is the magnetic field strength.
In the frame where $\mu=0$ for an electron,
a circular motion in the same plane is observed, and the radiation is symmetric with respect to the plane
so that the electron is kept in the same plane.
Therefore, even in the laboratory frame,
the radiation reaction does not change the velocity along the magnetic field $\beta \mu$,
where $\beta=\sqrt{1-1/\gamma^2}$.
Combining $\beta\mu=\mathrm{const.}$ and the equation (\ref{eq:synenergy}), we obtain 
the pitch angle changing rate by radiation reaction \citep{1985ApJ...299..987P,2016MNRAS.458.2303S} as
\begin{eqnarray}
\label{eq:synpitch}
\dot{\mu}_{\mathrm{syn}}=\frac{2e^4B^2\mu(1-\mu^2)}{3m_{\mathrm{e}}^3c^5\gamma}.
\end{eqnarray}

We assume a toroidal dominant magnetic field decreasing with the jet expansion as $B\propto R^{-1}$,
so we write
\begin{eqnarray}
\label{eq:B}
B=B_{0}\left(\frac{R_{0}}{R}\right)=B_{0}\left(1+\frac{t}{t_\mathrm{d}}\right)^{-1},
\end{eqnarray}
where $B_{0}$ is the magnetic field strength at $R=R_{0}$.
Using the time $t$ in the comoving frame, the radius increases as $R=R_{0}+\Gamma \beta_{\mathrm{j}} ct$, where the jet velocity $\beta_{\mathrm{j}}=\sqrt{1-1/\Gamma^2}$.
We have defined the dynamical timescale as 
\begin{eqnarray}
\label{eq:dynamicaltime}
t_\mathrm{d} \equiv \frac{R_{0}}{\Gamma\beta_{\mathrm{j}} c}.
\end{eqnarray}

Even if the field $B$ decreases, an electric field along the magnetic field is not induced
($\dot{\bm{B}}=-c \nabla \times \bm{E}$).
As the force along the field $B$ is zero, the parallel momentum $p_{\parallel}\equiv
m_{\mathrm{e}}c\gamma \beta\mu$ is conserved.
With the adiabatic invariant $p_{\perp}^2/B$ \citep{1963RvGSP...1..283N}, where the perpendicular momentum
$p_{\perp}\equiv m_{\mathrm{e}}c \gamma \beta\sqrt{1-\mu^2}$,
we obtain the adiabatic energy changing rate
\begin{eqnarray}
\label{eq:adienergy}
\dot{\gamma}_{\rm adi}=-\frac{(\gamma^2-1)(1-\mu^2)}{2\gamma}\left(t_{\mathrm{d}}+t\right)^{-1},
\end{eqnarray}
and the adiabatic pitch angle changing rate
\begin{eqnarray}
\label{eq:adipitch}
\dot{\mu}_{\rm adi}=\frac{\mu(1-\mu^2)}{2}\left(t_{\mathrm{d}}+t\right)^{-1}.
\end{eqnarray}

The evolution of the electron distribution is characterized by the ratio $t_{\mathrm{d}}/t_{\mathrm{c}}$,
where $t_{\mathrm{c}}$ is the initial timescale of the synchrotron cooling
for electrons of $\gamma=\gamma_{\mathrm{min}}$, 
\begin{eqnarray}
\label{eq:coolingtime}
t_{\mathrm{c}}=\frac{9 m_{\mathrm{e}}^3 c^5}{4 e^4 B_{0}^2 \gamma_{\mathrm{min}}},
\end{eqnarray}
where we adopt the average value in the isotropic distribution for the pitch angle  as $\langle\sin^2\psi\rangle=2/3$.

The ratio of the two pitch angle change rates is written as
\begin{eqnarray}
\label{eq:pitchratio}
\frac{\dot{\mu}_{\mathrm{syn}}}{\dot{\mu}_{\mathrm{adi}}}=\frac{3}{\gamma\gamma_{\mathrm{min}}}\frac{t_{\mathrm{d}}}{t_{\mathrm{c}}}\left(1+\frac{t}{t_{\mathrm{d}}}\right)^{-1}.
\end{eqnarray}
For $\gamma^2_{\mathrm{min}} \gg t_{\mathrm{d}}/t_{\mathrm{c}}$,
the pitch angle change due to synchrotron radiation is negligible.

\begin{figure}[htp]
  \centering
  \includegraphics[width=8cm]{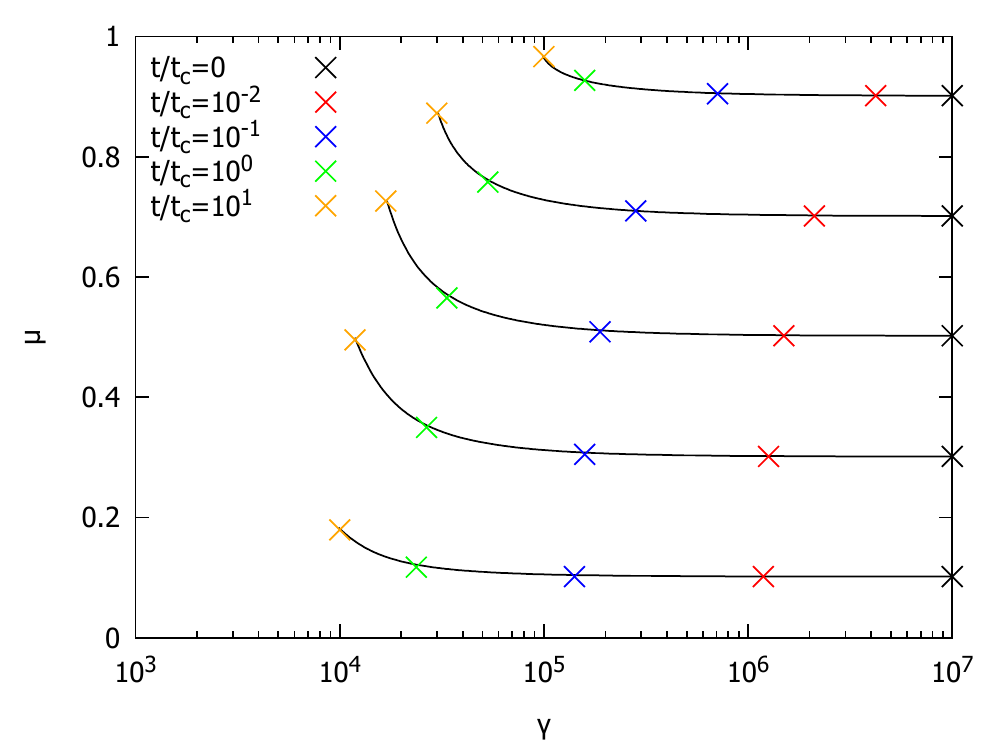}
  \caption{The temporal evolutions of electron trajectories in the $\gamma$-$\mu$ space in the case of $t_{\mathrm{d}}/t_{\mathrm{c}}=1$ with $\gamma_{\mathrm{min}}=10^4$.
  Electrons are injected with $\gamma=10^7$ and different values of $\mu$ at $t=0$.}
  \label{fig:gammamu}
\end{figure}

Figure \ref{fig:gammamu} shows electron trajectories in the $\gamma$-$\mu$ space, where the initial $\gamma$ is $10^7$ with various pitch angles.
Initially, the cooling timescale of electrons of $\gamma=10^7$ is $10^3$ times shorter
than that for $\gamma=\gamma_{\mathrm{min}}$,
so those electrons promptly lose their energy with negligible changes of $\mu$.
For $t> t_{d}$, the magnetic field decreases significantly and the synchrotron cooling is suppressed.
The electrons mainly lose their energies by adiabatic cooling.
From the equation (\ref{eq:adipitch}), the timescale of the pitch angle change by adiabatic cooling
is longer than $t$ for $t<t_{\mathrm{d}}$.
The pitch angle changes significantly for $t> t_{d}$.
From $\dot{\mu}_{\mathrm{adi}}\propto\mu(1-\mu^2)$, electrons with intermediate values of $\mu$
experience the change of the pitch angles most significantly.
Figure \ref{fig:gammamu} also shows that the synchrotron energy loss is relatively suppressed for smaller pitch angles (larger $\mu$).

\subsection{\bf Initial Pitch Angle Distribution} \label{sec:pitch}

Motivated by the simulations of turbulence magnetic reconnection in
\citet{2019ApJ...886..122C}, we assume anisotropic electron distributions at the injection.
According to \citet{2019ApJ...886..122C}, at the reconnection site
electrons are accelerated along the guiding magnetic field with small pitch angles,
and the succeeding turbulence acceleration gradually accelerates electrons perpendicular to the field.
The pitch angle distribution becomes isotropic at higher energies.
We adopt a Gaussian distribution in the $\mu$-space,
\begin{eqnarray}
\label{eq:pitchdistribution}
f(\gamma,\mu)=f_0(\gamma) \exp\left(-\frac{(\mu-\overline{\mu})^2}{2\Delta\mu^2}\right),
\end{eqnarray}
where $\Delta \mu$ is a function of $\gamma$, and the normalization $f_0(\gamma)$ is adjusted to satisfy
$\int_{-1}^{1} f(\gamma,\mu) d \mu=1$.
Most of low energy electrons have small pitch angles $\mu\simeq1$, so we adopt the peak pitch angle as
\begin{eqnarray}
\label{eq:mubar}
\overline{\mu}=1.
\end{eqnarray}
In the PIC simulations of \citet{2019ApJ...886..122C},
$\overline{\mu}$ becomes smaller at higher energies.
For simplicity, however, we fix $\overline{\mu}=1$ irrespectively of $\gamma$.
Alternatively, we assume a dispersion growing with $\gamma$ as
\begin{eqnarray}
\label{eq:deltamu}
\Delta\mu=\Delta\mu_{\rm{min}}\left(\frac{\gamma}{\gamma_{\rm{min}}}\right)^k,
\end{eqnarray}
where constant parameters $\Delta\mu_{\mathrm{min}}$ and $k$ are introduced.
As fiducial values, we adopt $\Delta\mu_{\mathrm{min}}=0.01$ and $k=1$.
In \citet{2019ApJ...886..122C}, the averaged $\sin\psi$ at minimum Lorentz factor $\gamma_{\mathrm{min}}$ is $\simeq0.14$,
which corresponds to $\Delta\mu_{\mathrm{min}}\simeq0.01$,
for the sigma parameter $\sigma\gtrsim10$ and the initial magnetic turbulent strength $\delta B/B=0.5$.
The pitch angle distribution becomes almost isotropic at $\gamma\simeq10\gamma_{\mathrm{min}}$ in the PIC simulations.
Those detailed results may depend on initial conditions and/or boundary conditions, so the value $k$ is not definite at present. In this paper, we adopt a value of $0.5<k<2$.

\begin{figure}[htp]
  \centering
  \includegraphics[width=8cm]{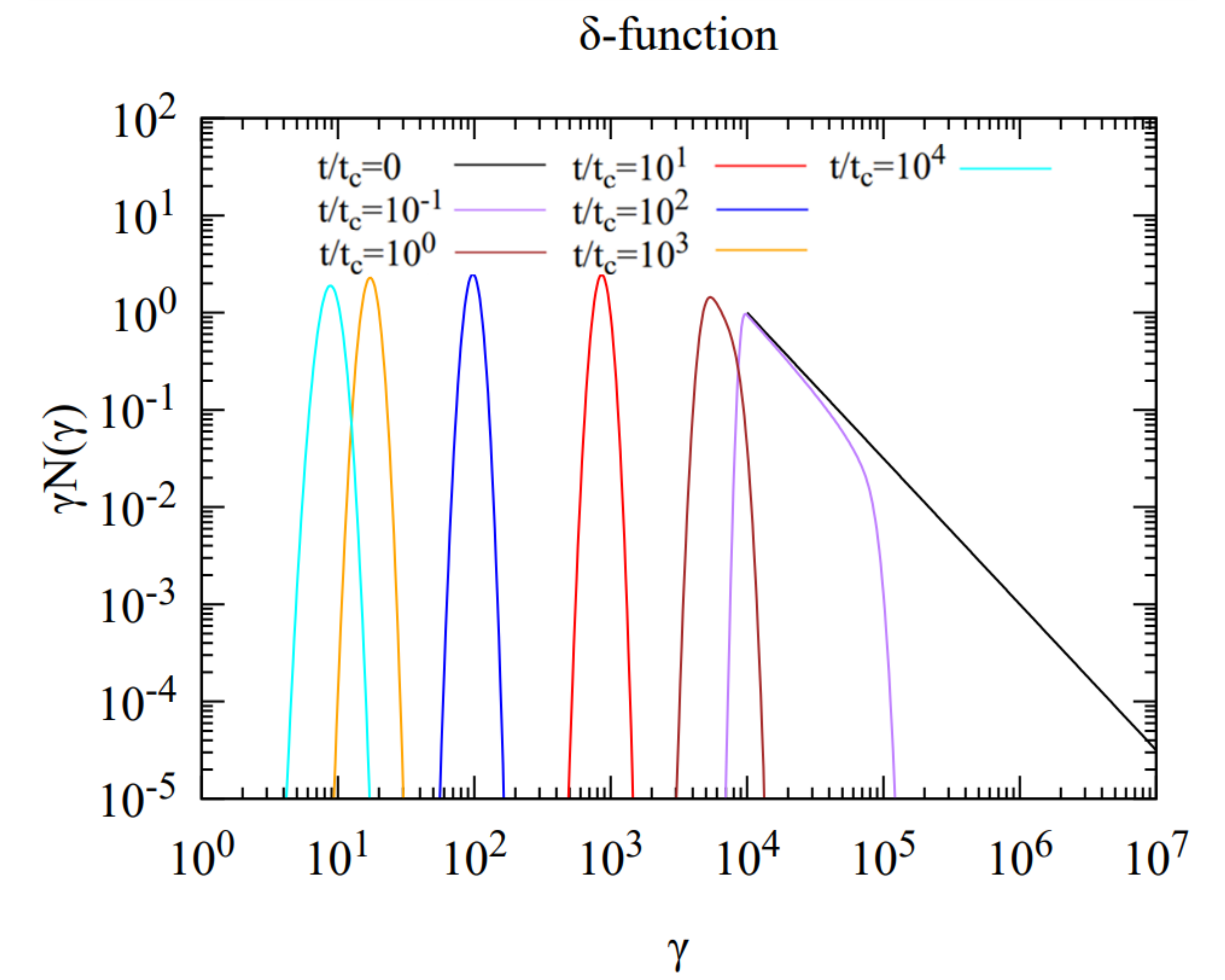}
  \includegraphics[width=8cm]{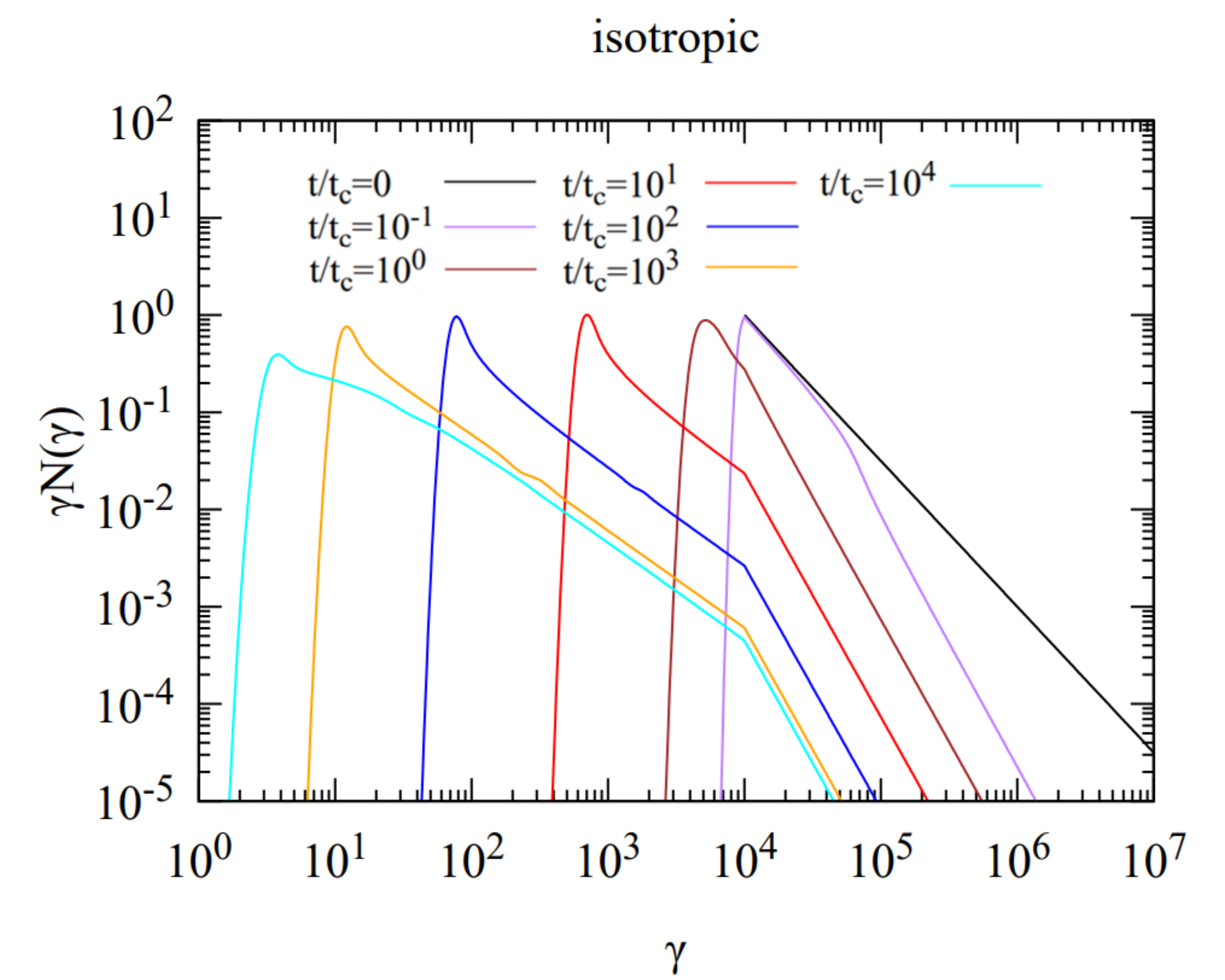}
  \includegraphics[width=8cm]{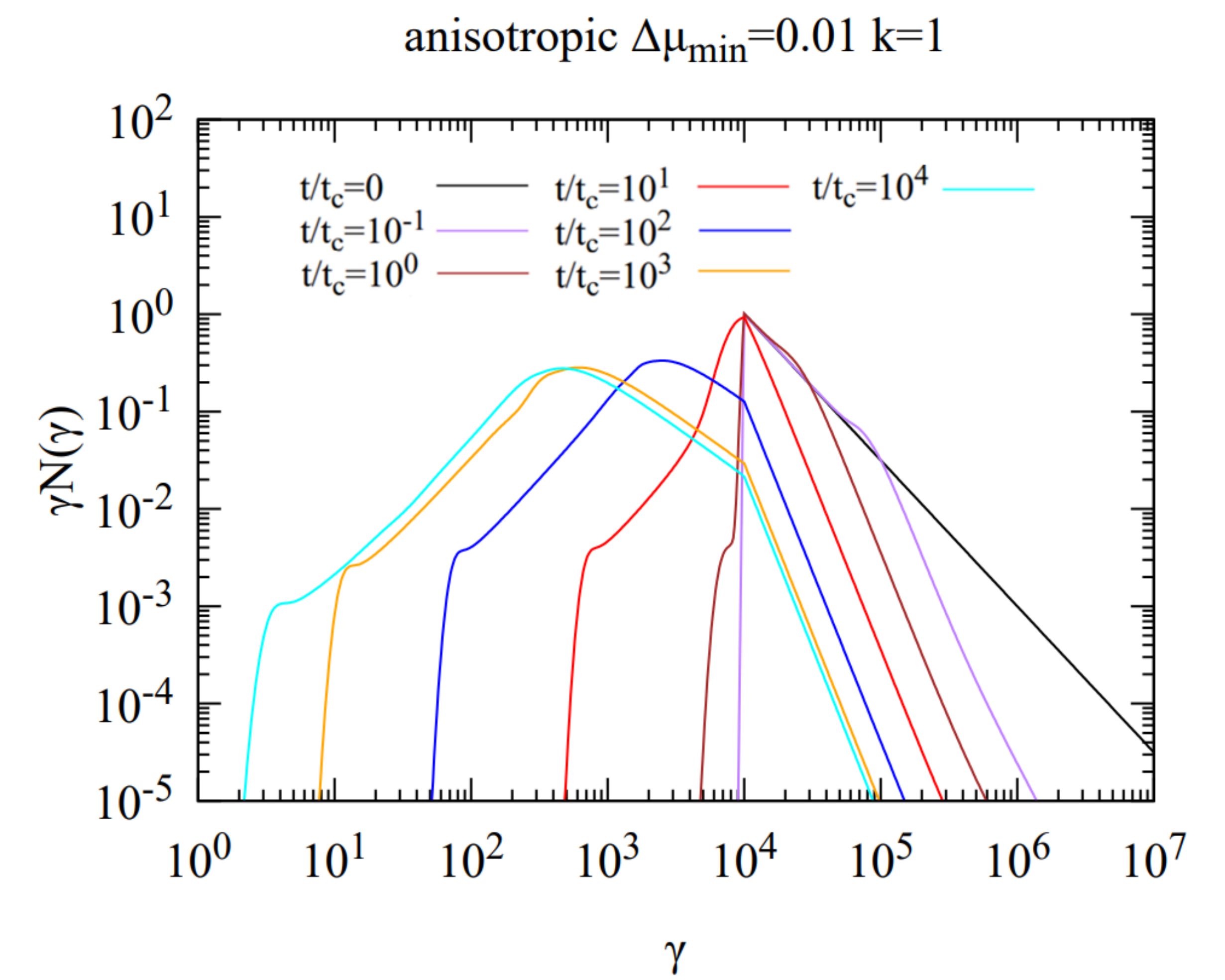}
  \caption{The evolutions of electron distributions for a single pitch angle
  ($f(\gamma,\mu)=\delta(\mu-\frac{1}{2})$, upper),
  isotopic distribution ($f(\gamma,\mu)=1/2$, middle)
  and our anisotropic model ($\Delta\mu_{\mathrm{min}}=0.01,k=1$, lower). The other parameters
  are common as $\gamma_{\mathrm{min}}=10^4$ and $t_{\mathrm{d}}/t_{\mathrm{c}}=10^3$.}
  \label{fig:cooling}
\end{figure}

In Figure \ref{fig:cooling}, we show the evolutions of the pitch-angle-integrated energy distributions of electrons: $\int_{-1}^{1}d\mu N(\gamma,\mu,t)\equiv N\left(\gamma,t \right)$.
We show three cases with different initial distributions: $f(\gamma,\mu)=\delta(\mu-\frac{1}{2})$ (single pitch angle),
$f(\gamma,\mu)=1/2$ (isotropic), and our anisotropic model with equation (\ref{eq:pitchdistribution}).
For the single pitch angle case, all the electrons have the same $\mu=0.5$, so they all cool down to the same Lorentz factor of $\gamma_{\mathrm{c}}(t)\simeq 9m_{\mathrm{e}}^3c^5/(4e^4B^2t)$.
For the isotropic case, the radiative cooling is suppressed only for electrons with very small pitch angles,
so such electrons remain in higher energy region even at a later stage.
This makes broken power-law energy distributions as shown in the middle panel.
After $t=t_{\mathrm{d}}=10^3t_{\mathrm{c}}$, the magnetic field decreases significantly.
At this stage, the cooling process switches from synchrotron cooling to adiabatic cooling.
For the anisotropic case of $\Delta\mu_{\mathrm{min}}=0.01,k=1$,
the cooling is further suppressed due to the small average of pitch angles compared to the isotropic case.

\subsection{\bf Synchrotron Spectrum}

We calculate time-integrated synchrotron spectra, because the observed spectra are usually provided as time-integrated one with an interval of a few seconds
due to poor photon statistics.
Integrating synchrotron radiation from the time $t=0$ to $t_{\rm end}$,
the spectrum is given by
\begin{eqnarray}
\label{eq:synspectrum}
\nu E_{\nu}=\int_0^{t_{\rm end}} dt \int_1^{\gamma_{\rm{max}}} d\gamma \int_{-1}^{1}d\mu N(\gamma,\mu,t)\nonumber\\
\times\frac{\sqrt{3}\nu e^{3}B\sqrt{1-\mu^2}}{m_{\mathrm{e}} c^2}F_{\mathrm{syn}}\left(\frac{\nu}{\nu_{\mathrm{syn}}}\right),
\end{eqnarray}
where $\nu$ is the photon frequency in the comoving frame, $E_{\nu}$ is the energy emitted per $\nu$, $\nu_{\rm{syn}} \equiv 3\gamma^{2}eB\sqrt{1-\mu^2}/(4\pi m_{e}c)$ is the synchrotron characteristic frequency,  $F_{\mathrm{syn}}\left(x\right) \equiv x \int_{x}^{\infty} K_{\frac{5}{3}}(\xi)d\xi$
and $K_{\frac{5}{3}}(\xi)$ is the modified bessel function of the $5/3$ order \citep{1979rpa..book.....R}.

In the following discussion, we focus on only the spectral shape, in particular on the photon spectral index $\alpha$. We define the spectrum $\langle{\nu E_{\nu}}\rangle$ that is normalized by the total energy of electrons,
\begin{eqnarray}
\label{eq:normalizedspectrum}
\langle{\nu E_{\nu}}\rangle\equiv \frac{\nu E_{\nu}}{E_{\mathrm{ini}}},
\end{eqnarray}
where
\begin{eqnarray}
\label{initialenergy}
E_{\mathrm{ini}}=\int_{\gamma_{\rm{min}}}^{\gamma_{\rm{max}}} d\gamma \int_{-1}^{1}d\mu \gamma m_{\mathrm{e}}c^{2}N(\gamma,\mu,t=0).
\end{eqnarray}
We shift the frequency in the jet's comoving frame to the observer frame
with $\nu_{\rm{obs}}=\Gamma \nu$.

\section{Results}  \label{sec:result}

\subsection{\bf Spectral Shape}

\begin{figure}[htp]
  \centering
  \includegraphics[width=8cm]{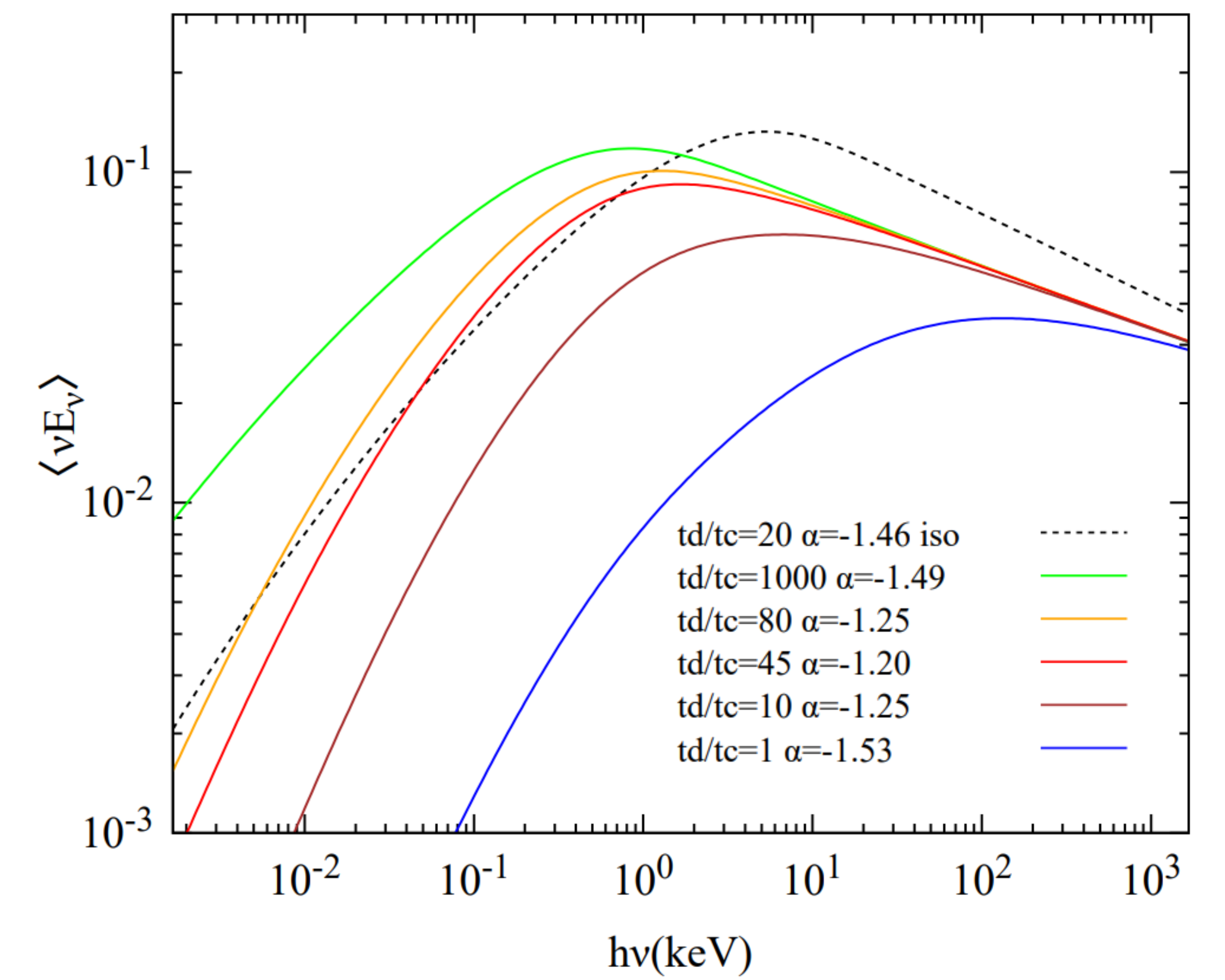}
  \caption{The calculated spectra in the comoving frame for $\gamma_{\mathrm{min}}=10^4$ and $B_{0}=2500\mathrm{G}$.
  The solid lines are spectra for the anisotropic model with $\Delta \mu_{\mathrm{min}}=0.01$, $k=1$,
  and $t_{\rm end}=10t_{\mathrm{d}}$ for different $t_{\mathrm{d}}/t_{\mathrm{c}}$ values.
  The dotted black line is an isotropic case with $t_{\mathrm{d}}/t_{\mathrm{c}}=20$ and $t_{\rm end}=10t_{\mathrm{d}}$.}
  \label{fig:spectra}
\end{figure}

Figure \ref{fig:spectra} shows the calculated spectra for various $t_{\mathrm{d}}/t_{\mathrm{c}}$ values.
Note again that the cooling time $t_{\mathrm{c}}$ is one for the isotropic distribution so that electrons above $\gamma_{\rm min}$ may not cool even for $t>t_{\rm c}$ depending on the degree of anisotropy.
In Figure \ref{fig:spectra}, we stop the calculation at $t_{\rm end}=10t_{\mathrm{d}}$,
at which the magnetic field decreases enough to suppress the synchrotron radiation.
In spite of the non-trivial anisotropic distributions,
those spectra can be well fitted with the Band function,
but the index $\alpha$ obtained from the fitting depends on the fitting energy range.
In this paper, we estimate the spectral index $\alpha$ at the photon energy of
$1/50$ times the peak energy of the $\nu E_{\nu}$ spectrum.
This energy corresponds to $\sim 10$ keV for the typical peak energy of $E_{\rm p}\sim 500$ keV.

In Figure \ref{fig:spectra},
the hardest spectral index is $\alpha \sim-1.2$ for $t_{d}/t_{c}=45$.
This large value of $t_{d}/t_{c}>1$ is due to suppression of cooling by anisotropic distribution with small pitch angles.
If we can perfectly block the electron cooling to $\gamma<\gamma_{\rm min}$,
the synchrotron emission should provide $\alpha=-0.67$.
but the radiation at $t>t_{\mathrm{d}}$ with the decreasing magnetic field
contributes to the low-energy region, and softens the spectrum as $\alpha \sim-1.2$.
For a much larger value of $t_{\rm{d}}/t_{\rm{c}}$, the index becomes $\alpha\sim-1.5$
(see the case for $t_{\mathrm{d}}/t_{\mathrm{c}}=1000$).
On the other hand, for a smaller value of $t_{d}/t_{c}$,
we again obtain a softer spectrum (see e.g. the case for $t_{\mathrm{d}}/t_{\mathrm{c}}=1$),
where the spectral peak is attributed to the cooling
break $\gamma_{\rm c}$ in the electron spectrum rather than $\gamma_{\rm min}$.
The low-energy spectrum in this case is attributed to the slow cooling electrons below $\gamma_{\rm c}$.

\begin{figure}[htp]
  \centering
  \includegraphics[width=8cm]{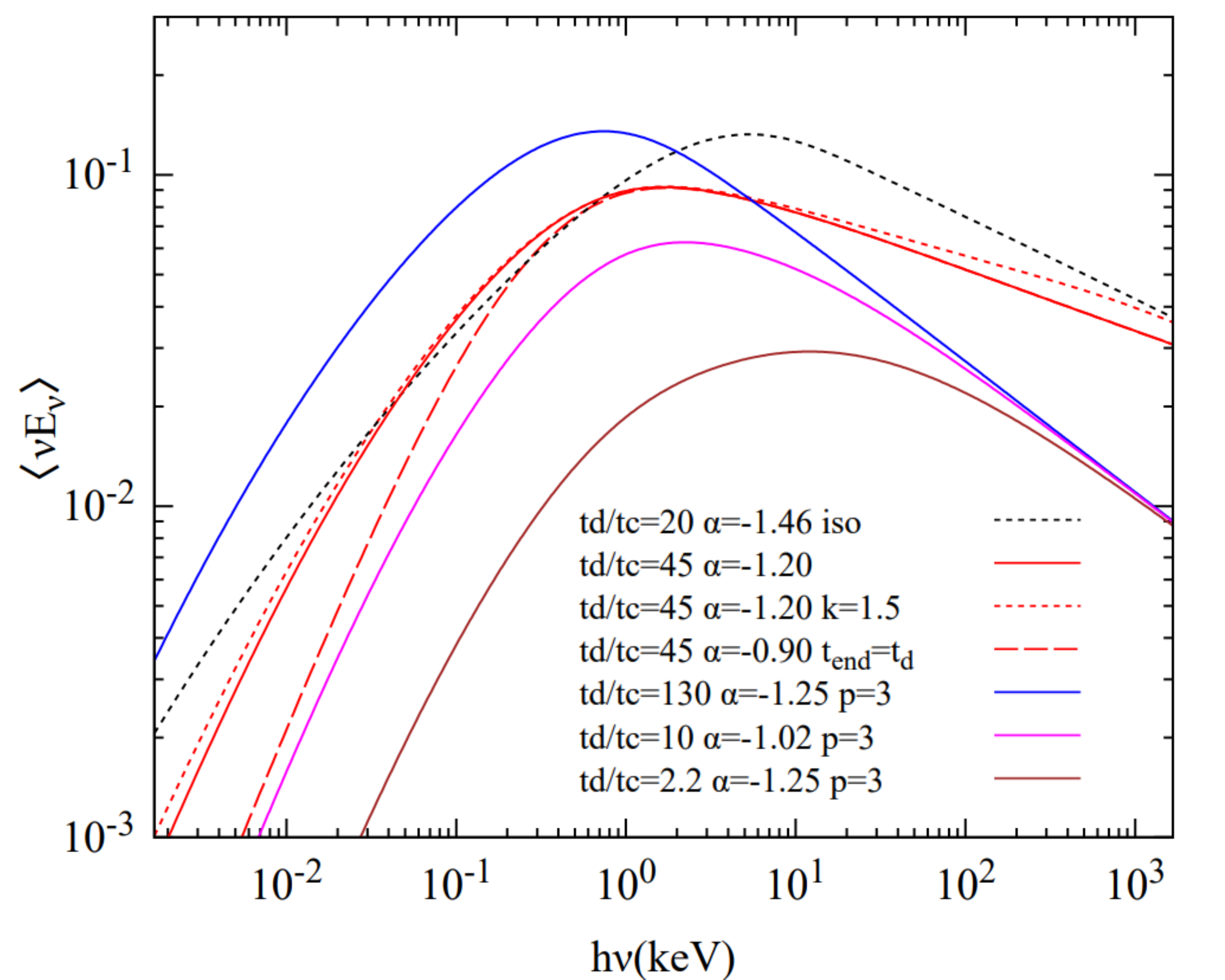}
  \caption{The calculated spectra in the comoving frame for $\gamma_{\mathrm{min}}=10^4$ and $B_{0}=2500\mathrm{G}$.
    The dotted black line is an isotropic case with $t_{\mathrm{d}}/t_{\mathrm{c}}=20$ and $t_{\rm end}=10t_{\mathrm{d}}$.
  The red lines are spectra for the anisotropic model with $\Delta \mu_{\mathrm{min}}=0.01$, changing the parameters as $t_{\rm end}=t_{\mathrm{d}}$ (dashed) and $k=1.5$ (dotted), respectively, from the reference model (solid, $k=1$ and $t_{\rm end}=10t_{\mathrm{d}}$), which is the same as the red line in Figure \ref{fig:spectra}.
  The blue, magenta and brown lines are the same anisotropic case with $p=3$ and different $t_{\mathrm{d}}/t_{\mathrm{c}}$ values. The other parameter values are the same as those for the reference case (red solid).}
  \label{fig:spectrap}
\end{figure}

A case with the isotropic distribution (the dotted black line)
shows a higher $E_{\rm p}$ than the anisotropic cases.
The lower peak energies in the anisotropic cases are due to the small pitch angles at $\gamma=\gamma_{\rm min}$,
as $E_{\rm p} \propto \sqrt{\langle \sin^2{\psi} \rangle} \sim \sqrt{2 \Delta \mu_{\rm min}}\sim 0.14$.
With decreasing $t_{\rm d}/t_{\rm c}$, the peak values of the $\nu E_{\nu}$
decreases, which shows the suppression of the radiation efficiency.
The peak energy $E_{\rm p}$ also increases with decreasing $t_{\rm d}/t_{\rm c}$
because of the suppression of electron cooling.
In Figure \ref{fig:spectra}, we have assumed $k=1$ for the solid lines,
whose flux above $E_{\rm p}$ is suppressed by the anisotropic distribution
compared to the isotropic case (the dotted black line).

Figure \ref{fig:spectrap} shows spectra for isotropic and anisotropic cases with different $p$, $k$ and $t_{\mathrm{end}}$.
When we stop calculation at $t=t_{\mathrm{d}}$ (dashed red), we obtain hard spectra with $\alpha=-0.90$,
because of the lack of the late radiation.
This optimistic assumption, the inefficient radiation for $t>t_{\mathrm{d}}$,
may be justified if the decay time of the shock-amplified magnetic field is comparable
to the dynamical timescale.
However, a kind of plasma instabilities responsible to the field amplification
implies a highly disturbed field, which may be incompatible with the negligible
pitch-angle diffusion that we have assumed.
Alternatively, the electron escape from the emission region with a strong magnetic field
can implement the termination of radiation at $t\sim t_{\mathrm{d}}$.

Another possible interpretation is that the temporal resolution of gamma-ray spectral observations is comparable to the dynamical timescale $t_{\rm d}/\Gamma$.
The typical time-bin for spectral analysis is a few seconds. So the variability timescale $R_0/(c \Gamma^2)$ should be comparable to a second in this interpretation. 

For $k=1.5$ (the dotted red line), the pitch-angle distribution rapidly approaches
an isotropic one with energy, so that the photon spectrum also
approaches to the flux of the isotropic model (the dotted black line)
at a higher energy.

For $p=3$ (blue, magenta and brown lines), the soft electron energy distribution yields radiation dominated by anisotropic electrons at the low energy region.
The contribution of cooled electrons isotropically injected at a higher energy to the low-energy flux is low compared to the cases with $p=2.5$.
The hardest spectral index for $p=3$ is $\alpha=-1.02$ with $t_{\mathrm{d}}/t_{\mathrm{c}}=10$ (magenta line), which is harder than $\alpha=-1.20$ in the $p=2.5$ case (red solid line).
The range of $t_{\mathrm{d}}/t_{\mathrm{c}}$ to make $\alpha$ harder than $-1.25$ also becomes wide as $t_{\mathrm{d}}/t_{\mathrm{c}}=2.2$-130 for $p=3$ case
($t_{\mathrm{d}}/t_{\mathrm{c}}=10$-80 for $p=2.5$).

As demonstrated in \citet{2014NatPh..10..351U}, we also test the continuous injection model, in which electrons with $f(\gamma,\mu)=\delta(\mu-\frac{1}{2})$ are injected continuously until $t=t_{\mathrm{d}}$. Even in this case, $t_{\mathrm{d}}/t_{\mathrm{c}}$
is required to be $\sim1$ to satisfy $\alpha>-1.25$.  As shown in \citet{2014NatPh..10..351U}, the instantaneous spectrum of emissivity at $t=t_{d}$ can be as hard as $\alpha=-1.07$ for $t_{\mathrm{d}}/t_{\mathrm{c}}=10$. However, the time-integrated spectrum yields $\alpha=-1.42$.

\subsection{\bf Allowed Parameter Regions}

\begin{figure}[htp]
  \centering
  \includegraphics[width=8cm]{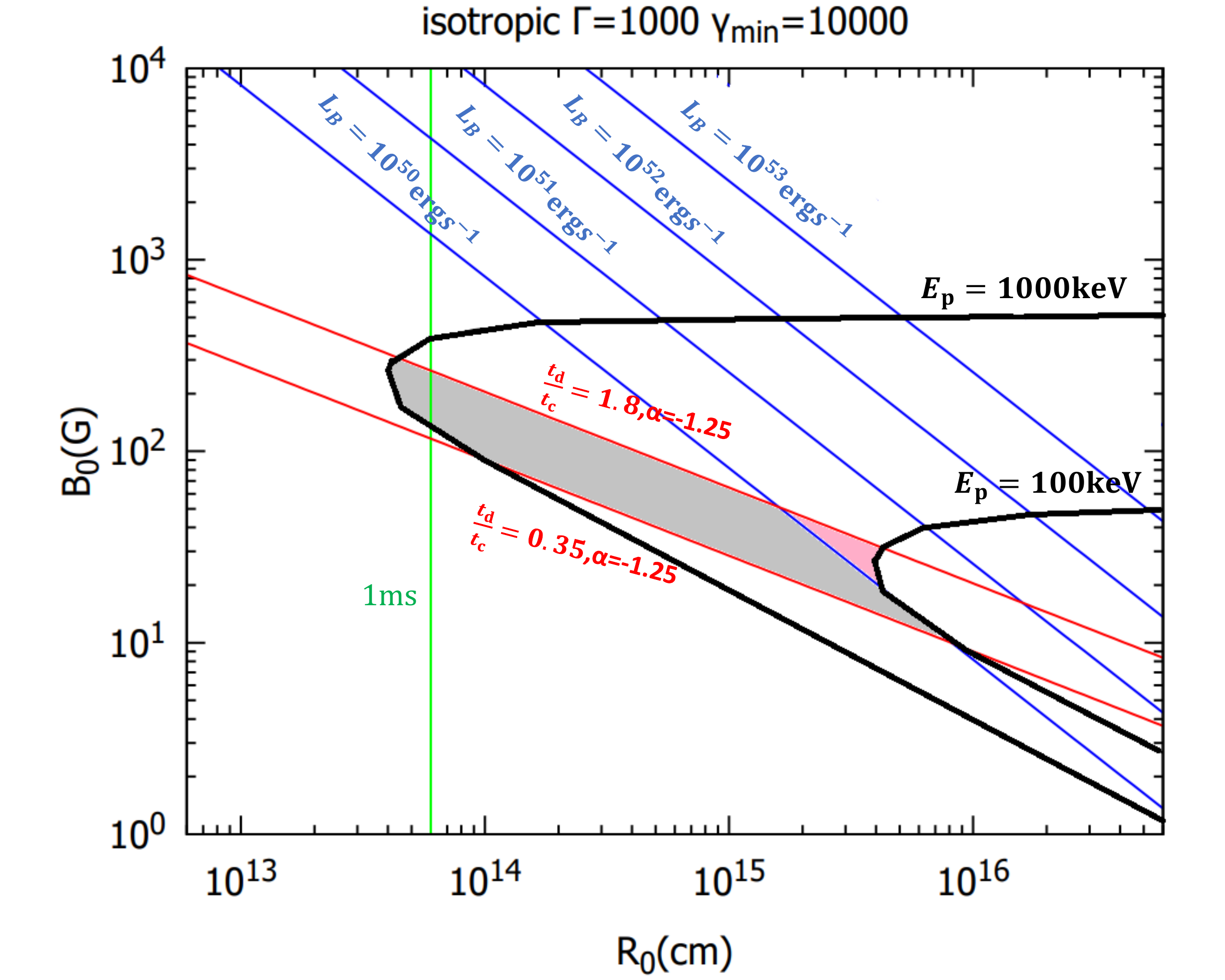}
  \includegraphics[width=8cm]{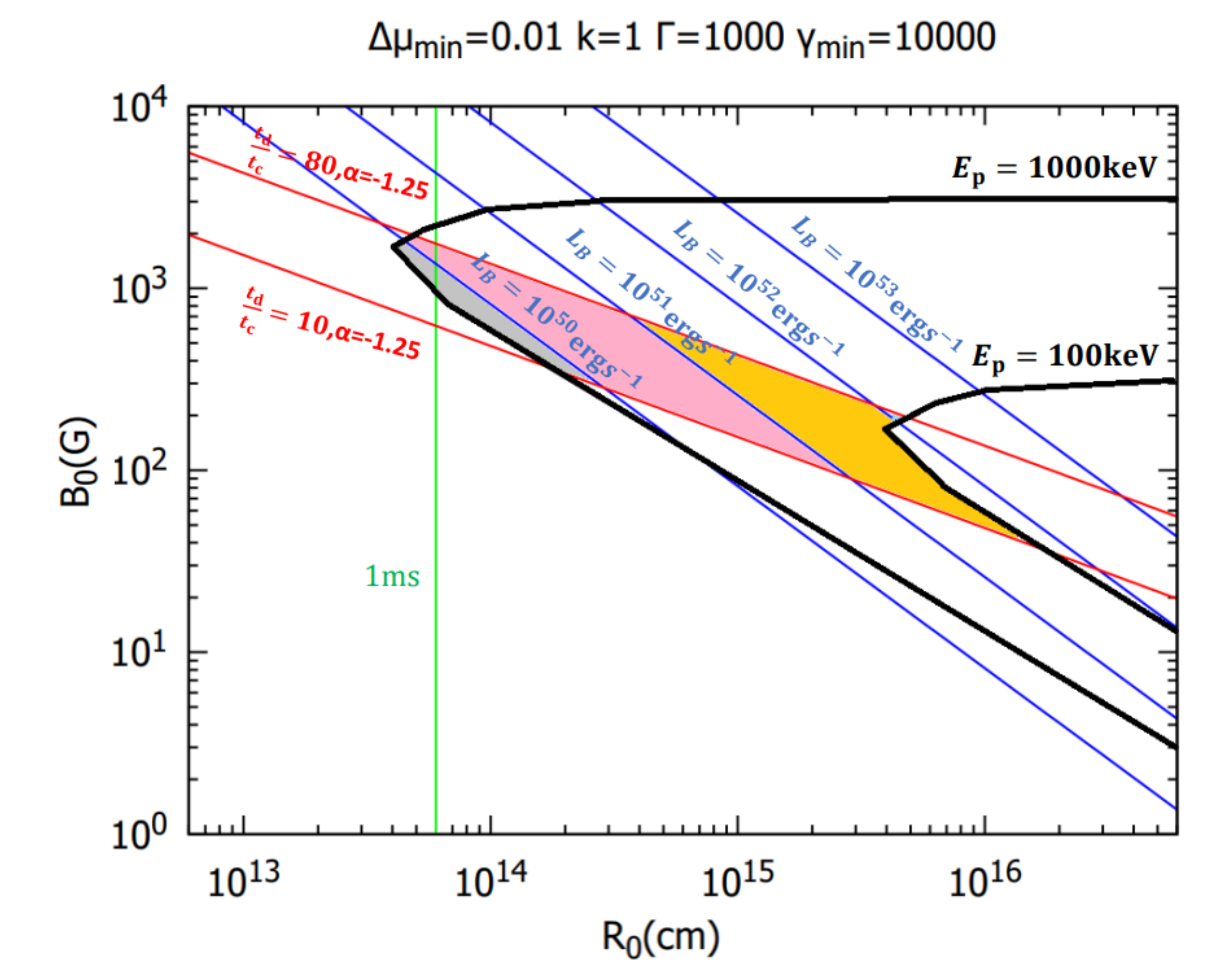}
  \includegraphics[width=8cm]{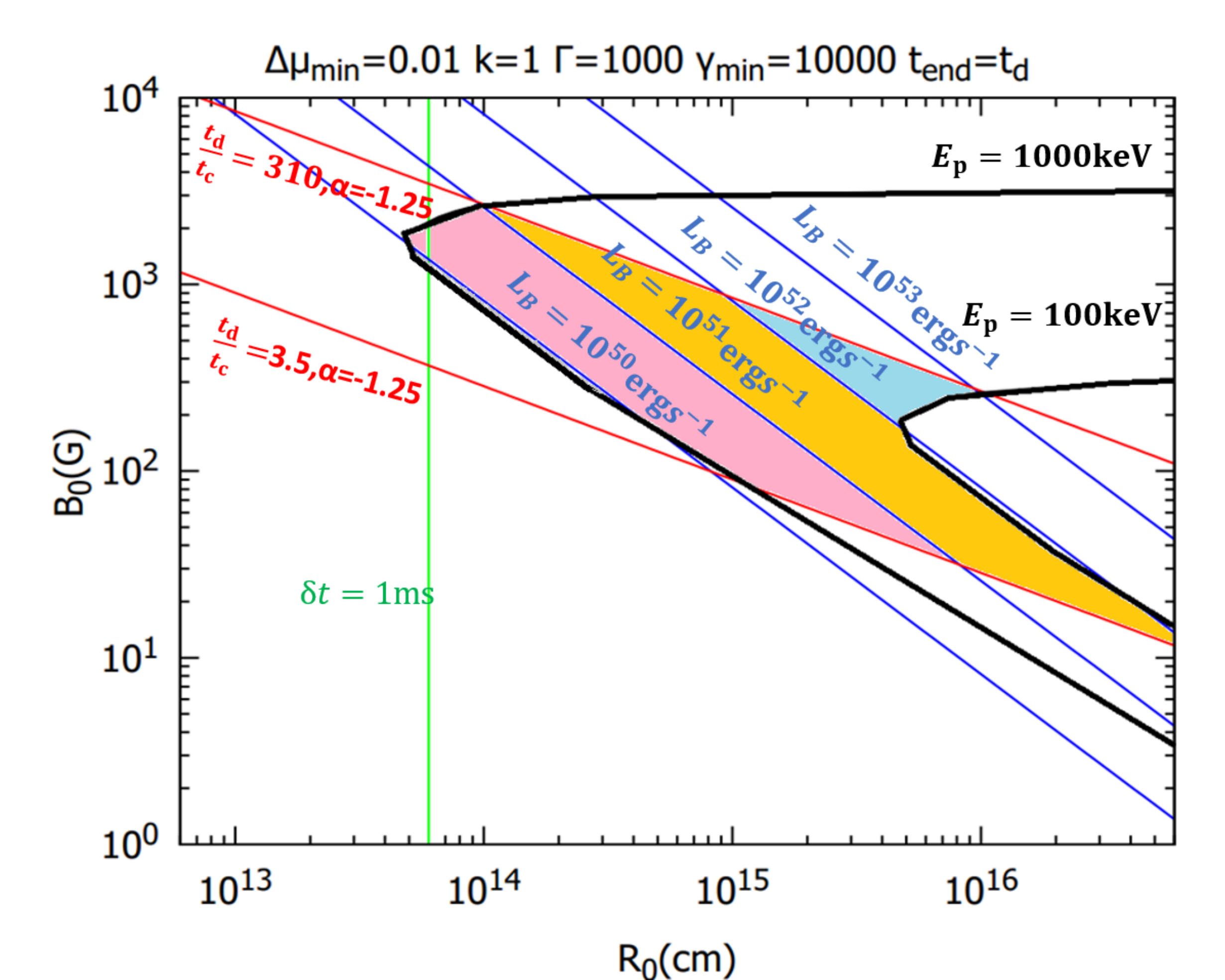}
  \caption{The allowed parameter regions for $\Gamma=1000$ and $\gamma_{\mathrm{min}}=10^4$
  are shown in the isotropic (top), anisotropic cases
  with $t_{\rm end}=10t_{\mathrm{d}}$ (middle) and $t_{\rm end}=t_{\mathrm{d}}$ (bottom).
In the anisotropic cases, the parameters are $\Delta \mu_{\mathrm{min}}=0.01$, and $k=1$ .
The two red lines enclose the region for $\alpha>-1.25$,
and the two black lines enclose the region for $100\mathrm{keV}<E_{\mathrm{p}}<1000\mathrm{keV}$.
The blue lines show constant magnetic luminosities.
In the allowed regions, $10^{50}\mathrm{erg/s}<L_{B}<10^{51}\mathrm{erg/s}$, $10^{51}\mathrm{erg/s}<L_{B}<10^{52}\mathrm{erg/s}$ and  $10^{52}\mathrm{erg/s}<L_{B}<10^{53}\mathrm{erg/s}$ are shown by magenta, orange and cyan areas, respectively.
The regions for $L_{B}<10^{50}\mathrm{erg/s}$ are shown by gray areas.
The green vertical line at $R_0=6\times 10^{13}\mathrm{cm}$ and the right edge $6\times 10^{16}\mathrm{cm}$ correspond to variability timescales $\delta t=R_0/(2 c \Gamma^2)=1\mathrm{ms}$ and $1\mathrm{s}$, respectively.}
  \label{fig:regiongamma1000}
\end{figure}

\begin{figure}[htp]
  \centering
  \includegraphics[width=8cm]{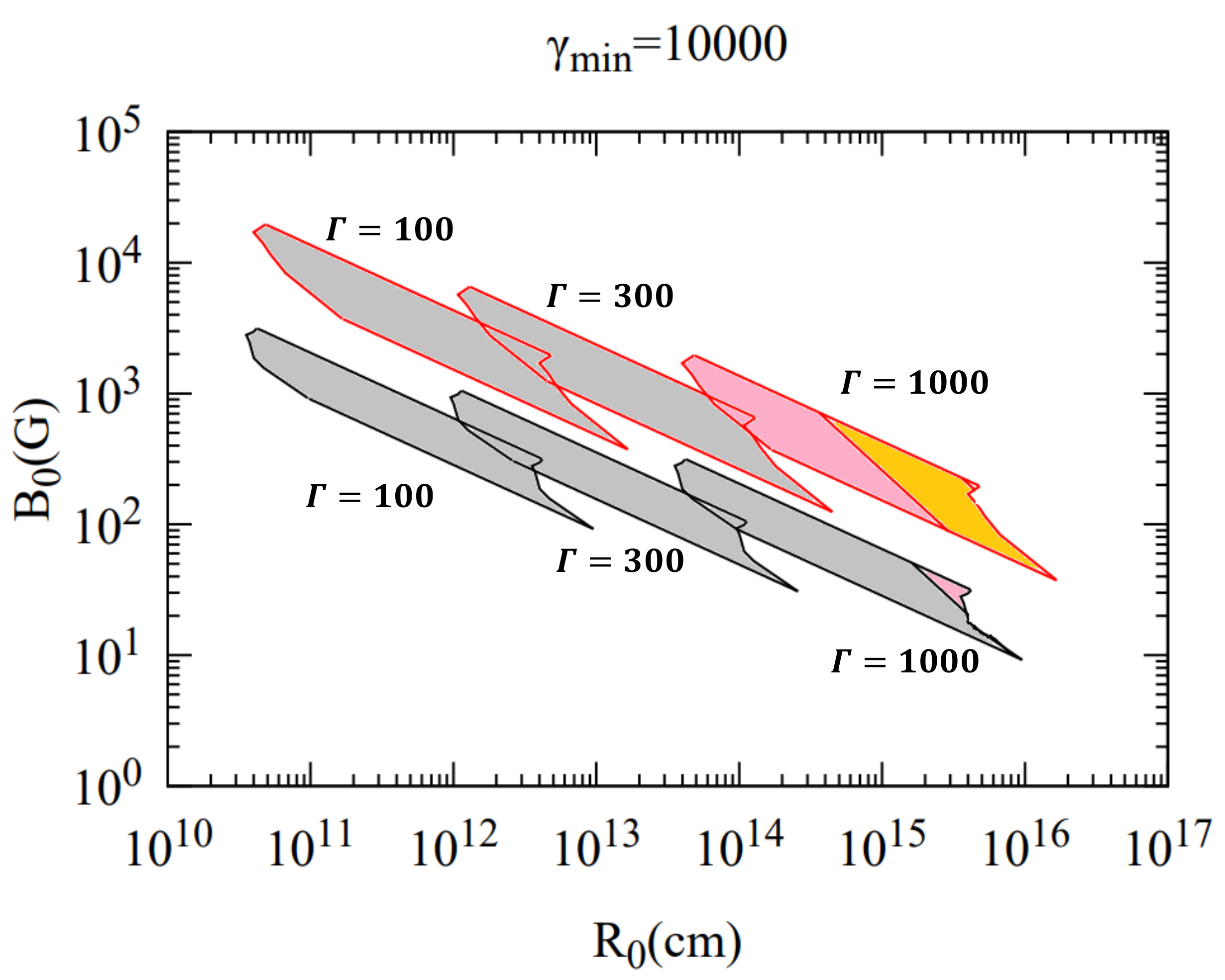}
  \includegraphics[width=8cm]{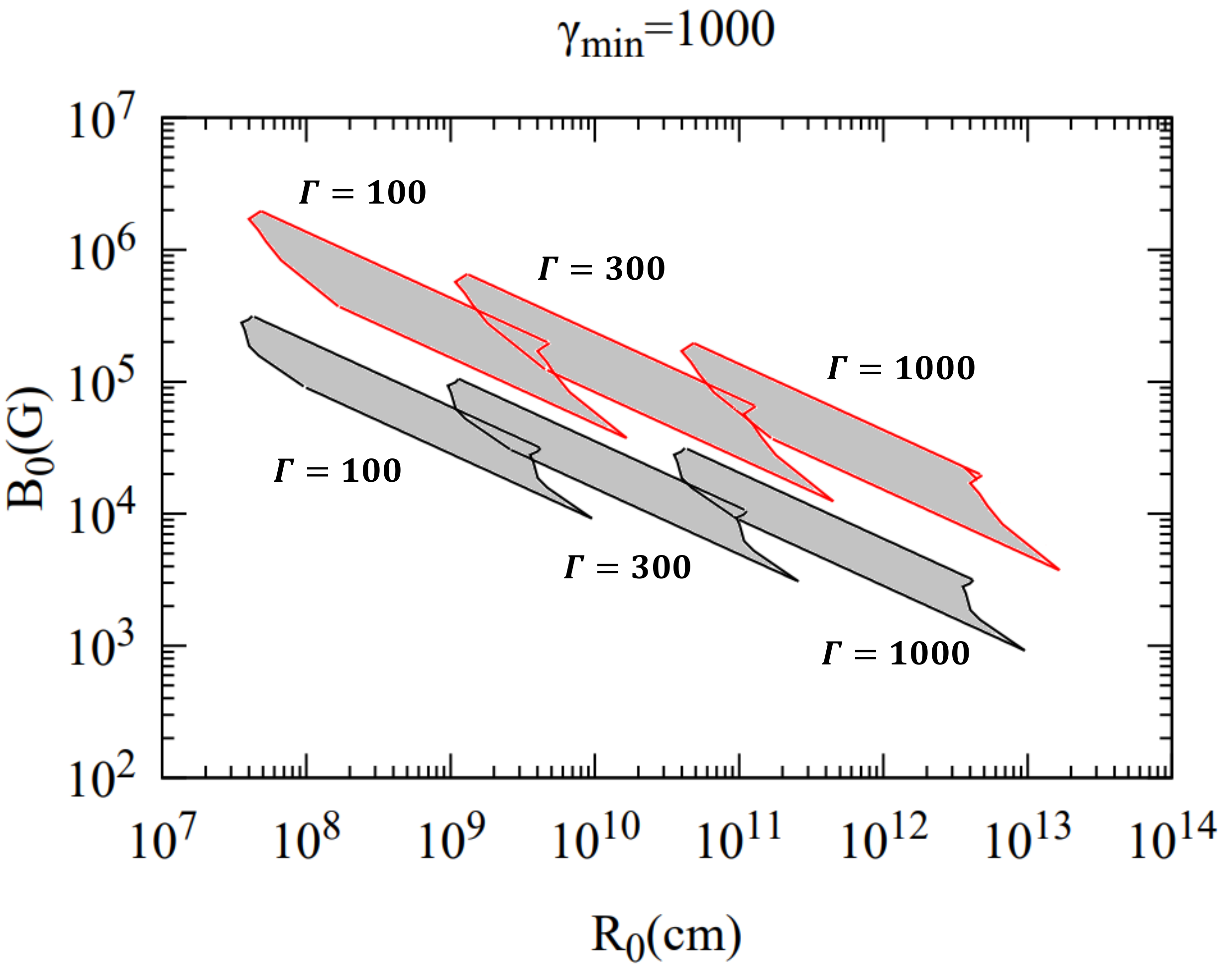}
  \caption{The allowed parameter regions are shown for the anisotropic case of $\Delta \mu_{\mathrm{min}}=0.01$ and $k=1$(red lines) and the isotropic case (black lines)
  for different values of $\Gamma$ and $\gamma_{\rm min}$ with $t_{\rm end}=10 t_{\mathrm{d}}$.
Upper panel: $\gamma_{\mathrm{min}}=10^4$ and Lower panel: $\gamma_{\mathrm{min}}=10^3$.
The magnetic luminosity is expressed with the same colors in Figure \ref{fig:regiongamma1000}.}
  \label{fig:region}
\end{figure}

Free parameters in our model are $\Gamma, \gamma_{\rm{\mathrm{min}}}, \Delta\mu_{\mathrm{min}},k, B_{0}$ and $R_{0}$.
We will show the allowed parameter regions in Figure \ref{fig:regiongamma1000} and \ref{fig:region} for the isotropic and anisotropic cases.
From the following two conditions to be consistent with the observed GRB properties,
we obtain allowed parameter regions of $B_{0}$ and $R_{0}$ for given values of $\Gamma, \gamma_{\rm{\mathrm{min}}}, \Delta\mu_{\mathrm{min}}$,
and $k$.

(I) Peak energy

In the standard synchrotron model,
the peak energy of the observed ${\nu_{\rm{obs}} F_{\nu_{\rm{obs}}}}$ spectrum is
determined by $\gamma_{\rm min}$ as,
\begin{eqnarray}
\label{eq:Ep}
E_{\mathrm{p}}=h\nu_{\mathrm{p}}\simeq\Gamma\frac{3 h \gamma_{\mathrm{min}}^2eB_{0}}{4\pi m_{\mathrm{e}} c}.
\end{eqnarray}
However, as shown in Figure \ref{fig:spectra}, the peak energy
increases with decreasing $t_{\mathrm{d}}/t_{\mathrm{c}}$.
Given $\gamma_{\mathrm{min}}$, $t_{\mathrm{d}}/t_{\mathrm{c}}$ and the other parameters
for the anisotropy,
the peak energy is proportional to $B_0 \Gamma$.
From the results numerically obtained,
we define the allowed region that satisfies $100\mathrm{keV}<E_{\mathrm{p}}<1000\mathrm{keV}$,

(I\hspace{-.1em}I) Low energy spectral index $\alpha$

The key parameter that determines $\alpha$ is $t_{\mathrm{d}}/t_{\mathrm{c}}$.
The maximum $\alpha$ is $\alpha\sim-1.2$ ($-1.0$) for $p=2.5$ ($3.0$) and $B\propto R^{-1}$.
Considering the observation uncertainty,
we require $\alpha>-1.25$ to reproduce the typical spectral index.

\begin{table}[htp]
  \caption{The lower and upper limits of $t_{\mathrm{d}}/t_{\mathrm{c}}$ that satisfy $\alpha>-1.25$ in the single $\mu$, isotropic and anisotropic cases with $t_{\rm end}=10t_{\mathrm{d}}$. The radiation efficiency $f_{\mathrm{rad}}$ is also shown in parentheses.}
  \label{}
  \centering
  \begin{tabular}{l|l|l}
  case & the lower limit & the upper limit \\ & of $t_{\mathrm{d}}/t_{\mathrm{c}}$ ($f_{\mathrm{rad}}$) & of $t_{\mathrm{d}}/t_{\mathrm{c}}$ ($f_{\mathrm{rad}}$) \\
  \hline\hline 
  single $\mu$ & 0.30 (0.44)& 1.7 (0.75)\\
    single $\mu$ steady injection & 0.36 (0.39) & 2.7 (0.71)  \\
  isotropic & 0.35 (0.43)& 1.8 (0.70)\\
  $k=0.5,\Delta\mu_{\rm{min}}=0.01$ & 10.5 (0.48) & 110 (0.80)\\
  $k=1,\Delta\mu_{\rm{min}}=0.01$ & 10 (0.55)& 80 (0.80)\\
  $k=1.5,\Delta\mu_{\rm{min}}=0.01$ & 12 (0.63)& 50 (0.79)\\
  $k=0.5,\Delta\mu_{\rm{min}}=0.1$ & 1.3 (0.47)& 10.5 (0.78)\\
  $k=1,\Delta\mu_{\rm{min}}=0.1$ & 1.7 (0.58)& 6 (0.75)\\
  $k=1.5,\Delta\mu_{\rm{min}}=0.1$ & 3.0 (0.69)& 3.5 (0.71)\\
  $k=1,\Delta\mu_{\rm{min}}=0.01,t_{\mathrm{end}}=t_{\mathrm{d}}$ & 3.5 (0.39) & 310 (0.90)  \\
  $k=1,\Delta\mu_{\rm{min}}=0.01,p=3$ & 2.2 (0.21) & 130 (0.79)  \\
  \end{tabular}
  \label{tab:tdtc}
\end{table}

In Table \ref{tab:tdtc}, we summarize the lower limit and upper limit of $t_{\mathrm{d}}/t_{\mathrm{c}}$
that satisfy $\alpha>-1.25$ for different cases.
The single-$\mu$ and isotropic cases require $t_{\mathrm{d}}/t_{\mathrm{c}} \sim 1$.
As the anisotropy strengthens, larger values of $t_{\rm d}$ compared to $t_{\rm c}$
are allowed.
Therefore, the fine tuning problem for $t_{\rm d}$ is relaxed in the anisotropic cases.

In Table \ref{tab:tdtc}, we also list the radiation efficiency, $\bm{f_{\mathrm{rad}}}$, defined as the energy ratio of the released photon energy to the initial total electron energy,
\begin{eqnarray}
&&f_{\mathrm{rad}} \nonumber \\  &&=\frac{\int_{0}^{\infty}d\nu E_{\nu}}{\int_{\gamma_{\rm{min}}}^{\gamma_{\rm{max}}} d\gamma \int_{-1}^{1}d\mu \gamma m_{\mathrm{e}}c^{2}N(\gamma,\mu,t=0)}.
\end{eqnarray}
The allowed parameter regions correspond to the marginally fast cooling, so that the radiation efficiency is significantly high as $\sim 0.2$--0.9.

In Figure \ref{fig:regiongamma1000}, we show examples of the allowed parameter regions
in the $R_0$-$B_0$ plane for several cases.
The lines that correspond to a constant value of $E_{\mathrm{p}}$ are curved.
In the fast cooling regime ($t_{\mathrm{d}}/t_{\mathrm{c}} \gg 1$),
the peak energy agrees with equation (\ref{eq:Ep}) so that the lines are horizontal
in Figure \ref{fig:regiongamma1000}.
On the other hand, in weak magnetic cases (slow cooling),
$E_{\mathrm{p}}$ increases with decreasing magnetic field as shown in Figure \ref{fig:spectra}.
This behaviour reflects on the curved feature of the constant $E_{\mathrm{p}}$ lines.

In Figure \ref{fig:regiongamma1000}, we also plot the magnetic luminosity,
\begin{eqnarray}
\label{eq:LB}
L_B=4\pi cR_0^2\Gamma^2\frac{B_{0}^2}{8\pi}.
\end{eqnarray}
If the anisotropic distribution is caused by magnetic reconnection,
the energy source of the non-thermal electrons should be the magnetic energy.
In this case, the magnetic luminosity should be larger than or
comparable to the observed gamma-ray luminosity $L_{\mathrm{\gamma,iso}}$
(note $f_{\mathrm{rad}} \sim 1$ even in our model).
So we present the high magnetic field regions with colored areas in Figure \ref{fig:regiongamma1000}.
As we adopt high values of $\Gamma$ and $\gamma_{\rm min}$ in Figure \ref{fig:regiongamma1000},
the allowed regions are compatible with significantly high $L_B$.
In the case of $t_{\rm end}=t_{\mathrm{d}}$, the allowed region is significantly wider
than the cases with $t_{\rm end}=10 t_{\mathrm{d}}$.

In Figure \ref{fig:region}, we show the allowed parameter regions in the $R_0$-$B_0$ plane for different values of $\Gamma$ and $\gamma_{\mathrm{min}}$.
If we require $L_{\rm{B}}>10^{50}\rm{ergs^{-1}}$ for the anisotropic model,
$\Gamma \gtrsim 1000$ and $\gamma_{\rm min} \sim 10^4$ are required.
From equations (\ref{eq:dynamicaltime}), (\ref{eq:coolingtime}), and (\ref{eq:Ep}), 
the initial radius is rewritten as 
\begin{eqnarray}
\label{eq:R}
R_{0}&=&10^{15}\left(\frac{t_{\rm d}/t_{\rm c}}{80}\right)\left(\frac{\Gamma}{1000}\right)^3\left(\frac{\gamma_{\mathrm{min}}}{10^4}\right)^3\nonumber\\&&\times\left(\frac{3E_{\mathrm{p}}}{300\mathrm{keV}}\right)^{-2}\mathrm{cm}.
\end{eqnarray}
The initial radius $R_{0}$ must be inside the deceleration radius of the afterglow, which is written as
\begin{eqnarray}
\label{eq:Rdec}
R_{\mathrm{dec}}&&=\left(\frac{3E_{\mathrm{iso}}}{4\pi n_{\mathrm{ex}}m_{\mathrm{p}}c^{2}\Gamma^{2}}\right)^{1/3}\nonumber\\
&&\simeq 10^{16}\left(\frac{E_{\mathrm{iso}}}{10^{52}\mathrm{erg}}\right)^{\frac{1}{3}}\left(\frac{n_{\mathrm{ex}}}{1\mathrm{\mathrm{cm^{-3}}}}\right)^{\frac{1}{3}}\nonumber\\&&\times\left(\frac{\Gamma}{1000}\right)^{-\frac{2}{3}}\mathrm{cm}.
\end{eqnarray}
where $m_{\mathrm{p}}$ is proton mass, $E_{\mathrm{iso}}$ is the total isotropic jet energy, $n_{\mathrm{ex}}$ is the ambient medium density.
If we take $\gamma_{\mathrm{min}}>10^5$ in equation (\ref{eq:R}), the allowed $R_0$
is larger than the deceleration radius of the afterglow around $10^{16}\mathrm{cm}$ \citep{2010ApJ...725.2209L}.
Therefore, only $\gamma_{\mathrm{min}}\sim 10^4$ is plausible for the anisotropic model
with magnetic reconnection.

For the isotropic models, a high value of $L_{\rm{B}}$ is not necessarily required.
In the standard internal shock model, in which the isotropic distribution is likely,
the typical value of $\gamma_{\rm min}$ is $\sim m_{\rm p}/m_{\rm e} \sim 10^3$
assuming that all electrons are accelerated.
As shown in the lower panel in Figure \ref{fig:region},
we find allowed parameter regions even for the isotropic distribution with $\gamma_{\rm min}=10^3$.
In this case, the magnetic luminosity is estimated to be lower than
$\sim10^{45}~\mbox{erg}~\mbox{s}^{-1}$ for $\Gamma=300$
or $\sim10^{48}~\mbox{erg}~\mbox{s}^{-1}$ for $\Gamma=1000$.
Such a lower magnetization may be favorable for particle acceleration
by relativistic shocks \citep[see e.g.][]{2020Galax...8...33V},
differently from the reconnection models.
However, the variability timescale $\delta t=R_0/(c \Gamma^2)$ should be shorter than
$\sim 0.03$ ms for $\Gamma=300$ or $0.1$ ms for $\Gamma=1000$, which could conflict
with the lower limit given by the light-crossing time for the Schwarzschild radius
of the central-engine black hole with a mass $M_{\rm BH}$,
$\sim 0.1 (M_{\rm BH}/10 M_\odot)$ ms.
The models with $\gamma_{\rm min}=10^4$ can avoid a too short $\delta t$
(shorter than 30 ms for $\Gamma=300$ or 100 ms for $\Gamma=1000$).
A large $\gamma_{\rm min}$ implies that a small fraction of electrons
are accelerated.

A too short $\delta t$ also leads to absorption of gamma-rays in
the source via electron--positron pair creation.
A simple estimate of the optical depth for $\gamma \gamma$-absorption
for photons of energy $E$
\citep{2003PASJ...55..433A,2009Sci...323.1688A} gives
\begin{eqnarray}
\tau_{\gamma \gamma}\simeq 0.1 \Gamma^{2 \beta-2}
\frac{(-\beta-2)\sigma_{\rm T}L_{\rm iso}}{16 \pi E_{\rm p}c^2 \delta t}
\left( \frac{E E_{\rm p}}{m_{\rm e}^2 c^4} \right)^{-\beta-1},
\end{eqnarray}
where $L_{\rm iso}$ is the isotropically equivalent luminosity.
To avoid absorption of MeV photons, a lower limit for the variability timescale
is given as
\begin{eqnarray}
\delta t > 0.9 \left(\frac{\Gamma}{100}\right)^{-6.4}
\left(\frac{L_{\rm iso}}{10^{52}\mbox{erg}~\mbox{s}^{-1}}\right)^{-1}
\left(\frac{E_{\rm p}}{100\mbox{keV}}\right)^{0.2}
\mbox{ms},\nonumber \\
\end{eqnarray}
where we adopt $\beta=-2.2$.
Only the isotropic model with $\Gamma=100$ and  $\gamma_{\rm min}=10^3$
in Figure \ref{fig:region} conflicts the above lower limit.

\section{Conclusions \& Discussion} \label{sec:sum}

In this paper, we have calculated the GRB synchrotron spectra with the anisotropic electron distribution,
which is motivated by the PIC simulations of magnetic reconnection in \citet{2019ApJ...886..122C}.
For that sake, we have followed the evolution of the electron distribution in the energy and pitch angle space
taking into account the synchrotron and adiabatic losses. 
Even with the anisotropic distribution,
the obtained spectra can be well fitted with the standard Band function.
Small pitch angles in the anisotropic electron distribution suppress the synchrotron cooling,
which makes synchrotron spectra harder.
Compared to the isotropic distribution, even with a stronger magnetic field,
the inefficient cooling can produce a significantly hard spectrum below the peak energy $E_{\rm p}$.
Given the parameters of $\Gamma$ and $\gamma_{\rm min}$,
allowed parameter regions of $R_0$ and $B_0$ becomes wider in the anisotropic model.

In our conservative assumption for the magnetic field evolution of $B\propto R^{-1}$,
the late radiation at $t>t_{\rm d}$ makes the low-energy spectral index $\alpha\sim-1.2$.
If we seriously consider the difference between the obtained index and the
typical observed index $\alpha\sim-1$,
a rapid decrease of the magnetic field or significant electron escape effect
after the dynamical timescale is required.

If the anisotropic electron injection is due to magnetic reconnection,
the magnetic luminosity should be comparable to or larger than the observed gamma-ray luminosity.
This condition requires the bulk Lorentz factor $\Gamma>1000$, because
the magnetic luminosity is expressed as
\begin{eqnarray}
\label{eq:LBpa}
L_{B}&=&2\times10^{51}\left(\frac{t_{\rm d}/t_{\rm c}}{80}\right)^2\left(\frac{\Gamma}{1000}\right)^6\left(\frac{\gamma_{\mathrm{min}}}{10^4}\right)^2\nonumber\\&&\times\left(\frac{3E_{\mathrm{p}}}{300\mathrm{keV}}\right)^{-2}~\mbox{erg}~\mbox{s}^{-1},
\end{eqnarray}
from equations (\ref{eq:dynamicaltime}), (\ref{eq:coolingtime}), (\ref{eq:Ep}), and (\ref{eq:LB}), where the factor $3$ before $E_{\mathrm{p}}$ is the difference of the peak energy of the anisotropic case from the isotropic fast cooling case (see Figure \ref{fig:spectra}). 
This also suggests that $\gamma_{\rm min} \gtrsim 10^4$ is simultaneously required.

There are allowed parameter regions for the isotropic electron distribution as well.
In this case, the standard shock acceleration model seems relevant.
The magnetic luminosity in the allowed region for $\Gamma=300$ and $\gamma_{\mathrm{min}}=10^4$ is around $10^{45}$--$10^{47}~\mbox{erg}~\mbox{s}^{-1}$,
which is consistent with the required low-magnetization for the efficient
particle acceleration by relativistic shocks \citep[e.g.][]{2013ApJ...771...54S,2020Galax...8...33V}.
However, to maintain the variability timescale long enough,
$\gamma_{\mathrm{min}}\gtrsim 10^4$ is required.

In this paper, we have neglected the effect of pitch angle diffusion by turbulence for simplicity.
The pitch angle scattering rate in the Alfv\'{e}n turbulence is given by 
\begin{eqnarray}
\label{eq:scatteringrate}
\nu_{\rm{sca}}=\frac{\pi}{4}\left(\frac{k_{\rm res} E(k_{\rm res})}{{B_{0}}^2/{8\pi}}\right)\Omega ,
\end{eqnarray}
where $E(k)$ is the power spectrum of the turbulence,
$k_{\rm res} \equiv \Omega/(c \mu)$ is the resonance wavenumber,
and $\Omega=eB_{0}/(\gamma m_{e}c)$ is the gyro frequency of an electron \citep{1987PhR...154....1B}.
Here, we assume a power spectrum of the Kolmogorov type turbulence as
\begin{eqnarray}
\label{eq:powerspectrum}
kE(k)\equiv\frac{{\delta B_{k}}^2}{8\pi}=\frac{{\delta B}^2}{8\pi}{\left(\frac{k}{k_{\mathrm{inj}}}\right)}^{-\frac{2}{3}},
\end{eqnarray}
where $\delta B$ is the the turbulent component of the magnetic field strength
at the injection wavenumber of the turbulence $k_{\mathrm{inj}}$,
for which we adopt the inverse of the comoving width of the emission region as
$k_{\mathrm{inj}}=\Gamma/R_{0}$,
which is the possible lowest wavenumber.
The condition to maintain the anisotropic distribution
is that the timescale of pitch angle scattering should be longer than $t_{\rm{d}}$;
\begin{eqnarray}
\label{eq:scatteringcondition}
\frac{1}{\nu_{\rm{sca}}}>\frac{R_{0}}{c \Gamma}.
\end{eqnarray}
This condition is written as 
\begin{eqnarray}
\label{eq:turbulencestrength}
\frac{\delta B}{B_{0}}&<&
{\left(\frac{4}{\pi}\right)}^{\frac{1}{2}}{\left(\frac{\Gamma r_{\mathrm{L}}}{R_0}\right)}^{\frac{1}{6}}{\left(\frac{1}{\mu}\right)}^{\frac{1}{3}}\nonumber\\
&\simeq&0.06\left(\frac{\Gamma}{1000}\right)^{\frac{1}{3}}\left(\frac{L_{B}}{10^{52} ~\mbox{erg}~\mbox{s}^{-1}}\right)^{-\frac{1}{12}}\nonumber\\&&\times\left(\frac{\gamma}{10^4}\right)^{\frac{1}{6}}{\left(\frac{1}{\mu}\right)}^{\frac{1}{3}},
\end{eqnarray}
where $r_{\mathrm{L}}=\gamma m_{e}c^2/(eB_{0})$ is the Larmor radius of an electron.
The obtained upper bound $\delta B/B_0 \sim 0.1$ seems reasonably large,
though the upper limit becomes smaller for a smaller injection scale,
which is determined by the reconnection layer.

Radiation from an anisotropic electron distribution could have a significant circular polarization degree.
When a circular polarization was detected in the optical afterglow of GRB 121024A, an anisotropic electron distribution was considered \citep{2014Natur.509..201W}.
The circular polarization degree for an anisotropic electron distribution $N(\gamma,\mu)\propto\gamma^{-p}f(\gamma,\mu)$ \citep{1969SvA....13..396S} is given by
\begin{eqnarray}
\label{eq:polarization}
P_{\mathrm{cir}}&=&\frac{1}{\gamma_{\mathrm{min}}}\frac{(2+p)\cot\theta-\sin\theta g(\gamma_{\mathrm{min}},\theta)}{p}\frac{p+1}{p+\frac{7}{3}}\nonumber\\& &\times\frac{\Gamma\left(\frac{3p+8}{12}\right)\Gamma\left(\frac{3p+4}{12}\right)}{\Gamma\left(\frac{3p+7}{12}\right)\Gamma\left(\frac{3p-1}{12}\right)},
\end{eqnarray}
where $\theta$ is the viewing angle from the magnetic field direction in the comoving frame, which is equivalent to the pitch angle of electrons that mostly contributes to radiation
for the observer.
Writing $\mu=\cos\theta$, the function $g$ is written as
\begin{eqnarray}
\label{eq:polarization2}
g(\gamma_{\mathrm{min}},\theta)=\left(\frac{1}{f(\gamma,\mu)}\frac{\partial f(\gamma,\mu)}{\partial \mu}\right)_{\gamma=\gamma_{\mathrm{min}}},
\end{eqnarray}
and $\Gamma$ in equation (\ref{eq:polarization}) is the Gamma function.
From order estimate for the anisotropic pitch angle distribution in our model with equation (\ref{eq:pitchdistribution}) using the typical viewing angle ($\theta \simeq 0.14$), we obtain $P_{\mathrm{cir}}\sim1/\gamma_{\mathrm{min}} \sim 0.01\%$ for $\Delta\mu_{\mathrm{min}}=0.01$ and $\gamma_{\mathrm{min}}=10^4$.
The predicted polarization degree is  only a few times larger than that for the isotropic case, which corresponds to $g(\gamma_{\rm min},\theta)\to 0$
in equation (\ref{eq:polarization2}).

\begin{acknowledgements}
First we appreciate the referee for the very helpful suggestions in spite of his/her difficult time.
We thank Kosuke Nishiwaki, Tomohisa Kawashima and Kyohei Kawaguchi for useful comments and discussion. R.G. received helpful advice for numerical calculation from Naotaka Yoshinaga and Chinatsu Watanabe and would like to thank them. 
R.G. acknowledges the support by the Forefront
Physics and Mathematics Program to Drive Transformation (FoPM).
This work is supported by the joint research program of the Institute for Cosmic Ray Research (ICRR), the University of Tokyo. 
Numerical computations were in part
carried out on FUJITSU Server PRIMERGY CX2550 M5 at ICRR. 

\end{acknowledgements}

\bibliography{ref}{}

\bibliographystyle{aasjournal}

\end{document}